\documentclass[amsmath,amssymb,aps,prd,showkeys,reprint]{revtex4-1}
\usepackage{graphicx}
\usepackage[breaklinks, colorlinks=true]{hyperref}
\usepackage{color}

\newcommand{\eqnref}[1]{Eq.~(\ref{#1})}
\newcommand{\secref}[1]{Sec.~\ref{#1}}
\newcommand{\appref}[1]{Appendix~\ref{#1}}
\newcommand{\tr}{\mathop{\mathrm{tr}}}
\newcommand{\p}{\partial}

\newcommand{\bbr}[1]{\left[#1\right]}
\newcommand{\cbr}[1]{\left\{ #1 \right\}}
\newcommand{\pbr}[1]{\left(#1\right)}
\newcommand{\Nc}{N_{\mathrm{c}}}

\newcommand{\fpi}{f_\pi}
\newcommand{\mpi}{m_\pi}

\newcommand{\EM}{\mathrm{em}}
\newcommand{\BA}{\mathrm{B}}
\newcommand{\bu}{\mathrm{bulk}}
\newcommand{\zm}{\mathrm{zm}}
\newcommand{\QB}{Q_{\BA,\bu}}
\newcommand{\QBZM}{Q_{\BA,\zm}}
\newcommand{\jchi}{j_{\EM,\chi}}
\newcommand{\jB}{j_{\BA,\bu}}
\newcommand{\JB}{\mathfrak{j}_{\BA}}
\newcommand{\jBZM}{j_{\BA,\zm}}

\newcommand{\phiA}{\varphi_{\mathrm{A}}}
\newcommand{\WZW}{\mathrm{WZW}}
\newcommand{\Lkin}{\mathcal{L}_{\chi}}
\newcommand{\Witten}{S_\WZW}
\newcommand{\axion}{a}
\newcommand{\spur}{\alpha}
\newcommand{\D}{\mathcal{D}}
\newcommand{\M}{\mathcal{M}} 
\newcommand{\MU}{{\bar\mu}}
\newcommand{\NU}{{\bar\nu}}
\newcommand{\RHO}{{\bar\rho}}
\newcommand{\SIGMA}{{\bar\sigma}}
\newcommand{\TAU}{{\bar\tau}}
\newcommand{\muB}{\mu_{\mathrm{B}}}

\begin{document}

\title{Anomaly inflow on QCD axial domain-walls and vortices}
\author{Kenji Fukushima}
\email{fuku@nt.phys.s.u-tokyo.ac.jp}
\author{Shota Imaki}
\email{imaki@nt.phys.s.u-tokyo.ac.jp}
\affiliation{Department of Physics, The University of Tokyo, 
  7-3-1 Hongo, Bunkyo-ku, Tokyo 113-0033, Japan}

\begin{abstract}
  We study the chiral effective theory in the presence of QCD
  vortices.  Gauge invariance requires novel terms from vortex
  singularities in the gauged Wess-Zumino-Witten action, which
  incorporate anomaly induced currents along the vortices.  We examine
  these terms for systems with QCD axial domain-walls bounded by
  vortices (vortons) under magnetic fields.  We discuss how the baryon
  and the electric charge conservations are satisfied in these systems
  through interplay between domain-walls and vortices, which manifests
  Callan-Harvey's mechanism of the anomaly inflow.
\end{abstract}
\maketitle

\section{Introduction}

Among various states realized on the phase diagram of matter out of
quarks and gluons, which are fundamental particles in quantum
chromodynamics (QCD), the region at high baryon density has been only
poorly understood and there are still diverging theoretical candidates
for the true ground state (see
Refs.~\cite{Rajagopal:2000wf,Alford:2007xm,Fukushima:2010bq,Fukushima:2013rx,Buballa:2014tba}
for comprehensive reviews over possible phases).  If the baryon
density is asymptotically high, we could perform the QCD-based
calculations to identify the ground state as the color-superconducting
state~\cite{Rapp:1997zu,Alford:1997zt,Alford:2003fq}.

We have learned high-temperature QCD matter from the relativistic
heavy-ion collision experiments, and such activities are going to be
continued toward new regimes with larger baryon density, which is
commonly called the beam energy scan program.  Besides, more and more
experimental data are expected from the neutron star observation which
will be a promising probe into dense QCD matter.  Interestingly, in
both cases of the heavy-ion collisions and the neutron star cores,
dense matter is exposed to strong magnetic fields.  As reviewed in
Ref.~\cite{Miransky:2015ava} many interesting and exotic phenomena,
mostly topological currents, are anticipated from the quantum anomaly
as result of the coupling between the density and the magnetic field
(see Ref.~\cite{Kharzeev:2015znc} for the implication to the heavy-ion
phenomenology).  In the context of the neutron star physics, in which
the magnetic field strength is of order $10^{13}$-$10^{15}\,\text{G}$
on the surface~\cite{harding2006physics} and stronger inside, magnetic
effects on color-superconducting quark matter have been
studied~\cite{Alford:1999pb,Ferrer:2005vd,Ferrer:2006vw,Ferrer:2006ie,Ferrer:2007iw,Noronha:2007wg,Fukushima:2007fc}.
Recently, on top of possible topological currents, an important aspect
of the chiral anomaly has been pointed out in
Ref.~\cite{Brauner:2016pko};  the ground state structure as the chiral
soliton lattice can be beautifully formulated for quark matter if the
magnetic field is sufficiently strong (see
Ref.~\cite{Fukushima:2013zga} for a related discussion on the
interplay between inhomogeneous quark matter and topological
currents).

Various QCD phases are characterized by various condensates leading to
the spontaneous breaking of global symmetries.  Then, according to the
broken symmetries, the physical vacuum has degeneracy and the
Nambu-Goldstone bosons are the dominant degrees of freedom for low
energy states.  For manifold which can wrap nontrivially over
configuration space, we can expect topologically stable
solitons~\cite{Manton:2004tk}.  Owing to the rich variety of QCD
phases, many different types of topological solitons are possible in
nuclear and quark matter~\cite{Eto:2013hoa}.  In this work we will
specifically consider the domain-wall and the vortex or vorton.

On the one hand, the domain-wall is a two-dimensional object that is
an interface between different energy minima.  Some theoretical
considerations hypothesize the presence of the $\pi^0$ domain-walls in
high density nuclear matter.  (It is also possible to think of
three-dimensional pionic profiles, which is exemplified in a scenario
of the Skyrme lattice~\cite{Forkel:1989wc}.)
Similarly, the $\eta$ domain-wall can stably exist in the color-flavor
locked (CFL) color-superconducting
phase~\cite{Son:2000fh,Son:2007ny}.  It is important to note that a
layered structure of the domain-walls is triggered by the anomaly
induced baryon charge on these axial domain-walls.  Thus, keeping the
total baryon charge, such a special configuration can reduce the
system energy.  We also mention another aspect of the anomaly;  an
anomaly induced axial current at finite
density~\cite{Metlitski:2005pr} and magnetic moment on axial
domain-walls~\cite{Son:2004tq} may cause spontaneous magnetization,
which may be a microscopic origin of the strong magnetic field on
magnetars (see also Refs.~\cite{Nakano:2004cd,*Tatsumi:2005ys,eto2013ferromagnetic,*hashimoto2015possibility}).

We note that, as argued in Ref.~\cite{Brauner:2016pko}, the
periodically layered structure of the axial domain-walls is a QCD
counterpart of
``chiral soliton lattice''~\cite{dzyaloshinskii1964theory,PhysRevLett.108.107202,%
kishine2015theory,togawa2013interlayer},
which refers to special spin aligned systems with breaking of parity
and translational symmetry.
The mathematical connection lies in the similarity between
the Dzyaloshinskii-Moriya interaction in chiral
materials~\cite{dzyaloshinsky1958thermodynamic,*moriya1960anisotropic}
and the Wess-Zumino-Witten (WZW) action in QCD.{}  This in turn means
that anomaly induced phenomena suggested in the QCD context may be
testable in analogous chiral materials.

The vortex is, on the other hand, a one-dimensional topological
defect with winding $\mathrm{U(1)}$ phase.  One example is the nuclear
vortex in superfluid nuclear matter, which could be continuously
connected to the CFL vortex associated with $\mathrm{U(1)_B}$
breaking~\cite{Forbes:2001gj}.  Another example is the $\eta$ vortex
in the CFL phase which spontaneously breaks $\mathrm{U(1)_A}$ by
diquark condensate.  Since these vortices carry angular
momenta~\cite{Davis:1989gn}, rotating CFL matter, as may be realized
in the neutron star cores~\cite{Alford:2000sx,Alford:2000sx}, would
form a lattice of these vortices.  Like the anomaly induced baryon
charge on the domain-walls, we can also anticipate anomaly induced
currents on the vortices such as an electric current along the $\eta$
vortex, an axial current along the baryon superfluid vortex,
etc~\cite{Metlitski:2005pr,Son:2004tq}.  Remarkably, such topological
currents are nondissipative and persistent, which is unusual in the
sense that the classical theorem to forbid any persistent current in
the ground state is evaded~\cite{Yamamoto:2015fxa}.

In this work we will consider a vortex ring.  Such a closed loop of
vortex string is often referred to as a vorton.  This special
configuration could emerge at the edge of the domain-wall disk or just
on its own.  Here, we make a historical remark about the idea of the
superconducting string~\cite{Witten:1984eb} that invoked serious
investigations on the cosmic string and possible vortons of the cosmic
string during the phase transitions in the Early
Universe~\cite{Kibble:1980mv,*Davis:1988ij,*Garaud:2013iba}.  Now, we
would emphasize that the ``QCD vorton'' could exist as a stable object
especially in the CFL phase with kaon
condensation~\cite{Buckley:2002ur}.  There is a nonzero charge and a
persistent superconducting current confined in the vortex
core~\cite{Kaplan:2001hh}.  Because this persistent current is
accompanied by finite angular momentum, the vorton is stabilized by the
angular momentum conservation~\cite{Buckley:2002mx,Buckley:2002mx}.
Moreover, a crucial difference of the CFL$K^0$ vorton from the cosmic
string is that the vorton forms a domain-wall due to the small mass of
the Nambu-Goldstone boson~\cite{Son:2001xd}.  This complements a
picture of the domain-wall disk surrounded by the vortex string, and
its shape is sometimes described as the ``drum vorton'' in the
literature~\cite{Buckley:2002ur,Carter:2002te}.

To investigate anomalous effects on QCD vortices, we will utilize the
chiral effective theory.  Because the low-energy dynamics is uniquely
determined by the symmetry breaking pattern, the formulation by the
chiral effective theory is quite robust and does not require any
microscopic details of nuclear matter or quark matter.  The
indispensable ingredient in theory is the WZW action to reproduce the
quantum anomaly with low-energy degrees of
freedom~\cite{Wess:1971yu,Witten:1983tw}.  The most well-known example
of the WZW action is the description of the $\pi^0\to\gamma\gamma$
decay process~\cite{Adler:1969gk, *Bell:1969ts} (see
Ref.~\cite{Nowak:2000wa} for this process at high density).  The full
construction of the WZW action is discussed in
Ref.~\cite{Kaiser:2000gs,*Kaiser:2000ck} and it has been applied to
the chiral magnetic effect in the hadronic
phase~\cite{Fukushima:2012fg}.

We will find in this work that, in the presence of vortex
singularities, the gauged WZW action requires new terms involving
derivative commutators; otherwise, gauge invariance is violated and
the charge conservation law is apparently broken.  These terms turn
out to incorporate anomalous currents absorbed or emitted by the
vortex, which is microscopically carried by zeromodes sitting on the
vortex.  

As a concrete physical setup, we will examine two different kinds of
finite-size axial domain-walls and vortices, namely, the $\pi^0$
vortex and the $\eta$ vortex, coupled to external magnetic fields.  We
will also argue that, with increasing magnetic fields, axial vortices
absorb or emit the baryon and the electric charges onto the
domain-wall.  Interestingly, such balance between the anomaly effect
on the domain-wall in bulk and that on the vortex at edge has a clear
interpretation as Callan-Harvey's mechanism of the anomaly
inflow~\cite{Callan:1984sa} once the axion vortex string is replaced
with the $\pi^0$ vortex or the $\eta$ vortex.

This paper is organized as follows:
In \secref{sec:anomaly}, we will derive the chiral effective theory
with the WZW action in the presence of the vortices.
In \secref{sec:axialdomainwall}, we will investigate two concrete
systems with the axial vortices and apply the fully gauge invariant
WZW action to see how the charge conservation is satisfied.
In \secref{sec:inflow}, we address the microscopic origin of the
vortex currents in terms of quark degrees of freedom along the
vortices.
The final section~\ref{sec:conclusion} is devoted to the conclusion.

\section{Topological current and conservation law}
\label{sec:anomaly}

We introduce the chiral effective theory and the WZW action.
In \secref{sec:cet} we extend the WZW action to a fully gauge
invariant form including vortex singularities.
In \secref{sec:conservation} we explicitly derive an extra
contribution to the topological current from the gauge invariant WZW
action and discuss how the conservation law is satisfied on axial
vortices.

\subsection{Chiral effective theory and currents}
\label{sec:cet}

The two-flavor chiral effective theory is characterized by the chiral
Lagrangian given in terms of triplet $\Sigma$ of pseudoscalar
Nambu-Goldstone bosons or pions $\pi_i$'s. 
The non-anomalous part of the leading order action
is, $S_\chi=\int d^4x\,\Lkin$, where
\begin{align}
	\Lkin
	&= \frac{\fpi^2}4 \tr\pbr{D_\mu\Sigma^\dagger D^\mu\Sigma 
	+ M \Sigma^\dagger + \Sigma M}
	\label{eq:Lchi}
\end{align}
with $D_\mu\Sigma = \p_\mu\Sigma + ieA_\mu[Q,\Sigma]$
and $Q=\mathrm{diag}(\frac23,-\frac13)$
in flavor space of $u$- and $d$-quarks%
~\cite{gasser1984chiral}.
The differentiation of this action with respect to the electromagnetic field $A_\mu$ 
yields an electric current,
\begin{align}
	\jchi^\mu
	= -\frac{\delta S_\chi}{\delta eA_\mu}
	= \frac{-i \fpi^2}2
		\tr\cbr{ \frac{\tau^3}2 \bbr{\Sigma D^\mu\Sigma^\dagger 
		- (D^\mu\Sigma^\dagger)\Sigma}}\,.
	\label{eq:jchi}
\end{align}
In the above expression we chose a convention to take the derivative
with $-eA_\mu$ for notational brevity.  In physics language
$\jchi^\mu$ represents the electric current associated
with $\pi^\pm$ flows.  Here, we shall introduce several notations which
we will frequently use in later discussions.  Vector fields made with
$\Sigma$ are commonly referred to as $L_\mu$ and $R_\mu$ defined by
\begin{align}
	L_\mu = \Sigma\partial_\mu \Sigma^\dagger\,,\qquad
	R_\mu = (\partial_\mu\Sigma^\dagger) \Sigma\,,
\end{align}
with which we can further construct second-order tensors
\begin{align}
	L_{\mu\nu} = \Sigma \partial_\mu\partial_\nu \Sigma^\dagger\,,\qquad
	R_{\mu\nu} = (\partial_\mu\partial_\nu\Sigma^\dagger) \Sigma\,.
\end{align}
We note that $L_{\mu\nu}$ and $R_{\mu\nu}$ are symmetric tensors if
the derivatives are commutative without vortex singularities.

Besides the non-anomalous current $\jchi$,
anomalous coupling between $\Sigma$ and external gauge fields
leads to the topological current induced by quantum anomaly.
This anomalous coupling is captured by the WZW action%
~\cite{Wess:1971yu, Witten:1983tw}.
The five-dimensional compact form of the WZW action is well-known as%
~\cite{Witten:1983tw}
\begin{align}
	\Witten^{(0)} 
		= \frac{i}{80\pi^2}\int_{\D} d^5 x\, \epsilon^{\MU\NU\RHO\SIGMA\TAU}
		\tr\pbr{L_{\MU}L_{\NU}L_{\RHO}L_{\SIGMA}L_{\TAU}} \,,
	\label{eq:witten0}
\end{align}
in the absence of gauge fields.
The indices with bar, $\MU,\,\NU,\,\RHO,\,\SIGMA,\,\TAU$, run
over five-dimensional spacetime denoted by $\D$.
The fifth coordinate
$x^4 \in [0,1]$ is chosen such that 
$\partial \D = \D(x^4 = 1)= \M$
where $\M$ stands for physical four-dimensional spacetime.

Anomaly induced electric current can be derived from the
action~\eqref{eq:witten0} as discussed below.
Under an infinitesimal $\mathrm U(1)_{\EM}$ rotation, i.e.,
$\Sigma \to \Sigma + i\zeta[Q,\Sigma]$ and
$A_\mu \to A_\mu - \partial_\mu\zeta/e$, the
action~\eqref{eq:witten0} varies up to the linear order in $\zeta$ as
\begin{align}
	\delta \Witten^{(0)} 
	= -\int_{\M} d^4 x\,(\p_{\mu}\zeta) X^\mu
	-\int_{\D} d^5 x\,(\p_{\MU}\zeta) Y^{\MU} \,.
\end{align}
We note that the above decomposition of $X^\mu$ and $Y^\MU$ is not unique.
The most convenient choice for our discussions is as follows;
\begin{align}
	X^\mu
		&= -\frac{1}{48\pi^2} \epsilon^{\mu\nu\rho\sigma}
		\tr(L_\nu L_\rho L_\sigma)\,,
	\label{eq:X} \\
	Y^{\MU}
		&= -\frac{1}{16\pi^2} \epsilon^{\MU\NU\RHO\SIGMA\TAU}
		\tr(L_{\NU}L_{\RHO}L_{\SIGMA\TAU})\,.
	\label{eq:Y}
\end{align}
We made this choice such that $Y^\MU$ would be vanishing for symmetric
$L_{\SIGMA\TAU}$.

We can modify the WZW action to make it gauge invariant up to the
linear order as
\begin{align}
	\begin{split}
	\Witten^{(1)} 
		&= \Witten^{(0)} 
			- \int_{\M} d^4x\, eA_\mu X^\mu
			- \int_{\D} d^4x\, eA_{\MU} Y^{\MU}\,.
		\label{eq:coupling}
	\end{split}
\end{align}
Yet, this action is not fully gauge invariant beyond the leading order.
We can reiterate the above procedures to the next order
to obtain $\Witten^{(2)}$ with corrections to $X^\mu$ and $Y^\MU$,
which is in fact fully gauge invariant.
Hereafter we will denote the fully gauge invariant WZW action by
$\Witten$.

In contrast, the baryon current cannot be obtained in the same way,
since $\Sigma$ is not sensitive to $\mathrm U(1)_\mathrm B$ rotation.
Instead, we identify the baryon current via Gell-Mann--Nishijima formula,
$Q = I_3 + B/2$%
~\cite{Nakano:1953zz, *Gell-Mann:1956iqa},
where $I_3$ and $B$ represent the isospin and the baryon number,
respectively.
The isospin part corresponds to nonanomalous $\jchi$
in \eqnref{eq:jchi}, whereas $B$ arises from anomaly within the
framework of the chiral effective theory.  Accordingly, we should
equate $B/2$ and the anomalous part in $Q$.
Further imposing the baryon conservation law%
~\cite{Vafa:1983tf},
we can construct the gauged WZW action as
\begin{align}
	\begin{split}
		\Witten = \Witten^{(0)}
		 & -\int_{\M} d^4 x\,\pbr{\spur_\mu+\frac{e}{2}A_\mu} \jB^\mu \\
		 & -\int_{\D} d^5 x\,\pbr{\spur_\MU+\frac{e}{2}A_\MU} \JB^\MU\,.
	\end{split}
	\label{eq:wzw}
\end{align}
Here a spurious gauge field $\spur_\mu = (\muB,\vec 0)$ implements the
coupling of the baryon chemical potential%
~\cite{Kogut:1999iv, Splittorff:2000mm, Son:2002zn}
with the topological
currents given by
\begin{equation}
	\begin{split}
		&\jB^\mu
		= -\frac{1}{24\pi^2}\epsilon^{\mu\nu\rho\sigma} \\
		&\times \Bigl\{ \tr(L_\nu L_\rho L_\sigma)
		- 3 i e (\p_\nu A_\rho) \tr[Q(L_\sigma+R_\sigma)] \\
		&- 3 i e A_\nu \tr[Q(L_\rho L_\sigma - R_\rho R_\sigma)] \Bigr\}
	\end{split}
	\label{eq:jB}
\end{equation}
in the four-dimensional part of the action and
\begin{equation}
	\begin{split}
		&\JB^{\MU}
		= -\frac1{8\pi^2}\epsilon^{\MU\NU\RHO\SIGMA\TAU} \\
		& \times \Bigl\{ \tr(L_{\NU}L_{\RHO}L_{\SIGMA\TAU})
		- i e (\p_{\NU}A_{\RHO})
		\tr[Q(L_{\SIGMA\TAU} + R_{\SIGMA\TAU})] \\
		& - i e A_{\NU}
		\tr[Q(L_{\RHO}L_{\SIGMA\TAU} \!-\! L_{\SIGMA\TAU}L_{\RHO}
	\!-\! R_{\RHO}R_{\SIGMA\TAU} \!+\! R_{\SIGMA\TAU}R_{\RHO})] \Big\}
\end{split}
\label{eq:JB}
\end{equation}
in the five-dimensional part.
We denote the ordinary four-dimensional current as $\jB$, for it
represents the current flowing not on vortices but in bulk. 
Here the first terms in
Eqs.~\eqref{eq:jB} and \eqref{eq:JB} come from
$X^\mu$ in \eqnref{eq:X} 
and $Y^\MU$ in \eqnref{eq:Y} 
in the leading order.
The four-dimensional current $\jB^\mu$ is 
well-known 
which traces back to Ref.~\cite{Wess:1971yu},
whereas the five-dimensional $\JB^\MU$ is a novel contribution in this work.
One can immediately see that 
the five-dimensional current $\JB^\MU$ would vanish
unless $\Sigma$ has a singularity, i.e.\
$[\partial_\MU, \partial_\NU]\Sigma^\dag \neq 0$~\cite{Kalaydzhyan:2014bfa}.

Finally let us define the total baryon charge out of these currents.
The baryon charge inferred from the ordinary four-dimensional current is
\begin{align}
	\QB = \int d^3\vec x\, \jB^0\,,
	\label{eq:QB}
\end{align}
and another charge contribution from the five-dimensional
current is
\begin{align}
	\QBZM = \int d^3\vec x \int_0^1 d x_4\, \JB^0\,.
	\label{eq:QBZM}
\end{align}
As we will discuss later, $\QBZM$ appears from the zeromode
contribution along vortices.
By construction, the anomaly induced electric charge
is given by the half of the baryon charge.
In this paper, therefore, we will focus on the baryon conservation only,
from which the electric charge conservation naturally follows.

\subsection{Charge conservation}
\label{sec:conservation}

If there is no singularity in $\Sigma$,
the current conservation is satisfied in a simple way.
That is, we see $\JB^\MU=0$ 
in a reason mentioned right below \eqnref{eq:JB} and
\begin{equation}
	\begin{split}
		&\p_\mu\jB^\mu
		= -\frac{1}{8\pi^2} \epsilon^{\mu\nu\rho\sigma} \\
		& \times \Bigl\{ \tr(L_\mu L_\nu L_{\rho\sigma} )
		- i e(\p_\mu A_\nu) \tr[Q(L_{\rho\sigma} + R_{\rho\sigma})] \\
		& - i eA_\mu \tr[Q(
		L_\nu L_{\rho\sigma} \!-\! L_{\rho\sigma} L_\nu
			\!-\! R_\nu R_{\rho\sigma} \!+\! R_{\rho\sigma} R_\nu )] \Bigr\}\,,
		\end{split}    
		\label{eq:divergenceofjB}
\end{equation}
which vanishes identically.
Therefore, $\QB$ by itself is conserved.

The situation would be, however, far more complicated when $\Sigma$
has a singularity leading to $[\p_\mu,\p_\nu]\Sigma^\dag\neq 0$.  Such
a singularity is typically associated with topological winding
of vortex configurations.
In the presence of a vortex flux
$[\p_\mu,\p_\nu]\Sigma^\dag$ is nonzero
proportional to the Dirac delta function at the vortex position.
Then, we must conclude,
\begin{align}
	\p_\mu \jB^\mu \neq 0\,,
\end{align}
and $\QB$ alone is no longer a conserved charge.

Let us see how this naive violation of the charge conservation law is
cured by the five-dimensional contribution.  By construction, the WZW
action in \eqnref{eq:wzw} possesses local gauge symmetry in $\D$.
Then, we can adapt gauge rotations in $\D$ not to affect $\M$, which
leads to
\begin{align}
	\p_{\MU}\JB^{\MU}(x,x^4) = 0\,.
	\label{eq:conservation1}
\end{align}
Conversely, if we apply gauge rotations on $\partial\D=\M$ at $x^4=1$,
we find,
\begin{align}
	\p_\mu\jB^\mu(x) = \JB^4(x,x^4=1)\,.
	\label{eq:conservation2}
\end{align}
We can give a plain interpretation for \eqnref{eq:conservation2}
that the current $\JB^4$ injected from $\D$ provides a source of
the baryon charge in $\M$. 
We can rewrite
\eqnref{eq:conservation2} in a form of the current conservation by
introducing the zeromode current, $\jBZM^\mu$, as
\begin{align}
	\jBZM^{\mu}(x) 
	&= \int_0^1 d x^4\, \JB^{\mu}(x,x^4)\,.
	\label{eq:jBZM}
\end{align}
We will discuss the meaning of the ``zeromode'' later in
\secref{sec:inflow}.
Then, using \eqnref{eq:conservation1},
we can rewrite the source term as
\begin{align}
	\JB^4(x^4=1)
	= \int_0^1 d x^4\, \p_4 \JB^4
	= -\p_\mu \jBZM^\mu \,.
        \label{eq:divjBZM}
\end{align}
We note that $\JB^4(x^4=0)=0$ follows from \eqnref{eq:conservation1}.
Finally we arrive at the baryon charge conservation law in the
following concise form,
\begin{align}
	\p_\mu (\jB^\mu + \jBZM^\mu) = 0\,.
	\label{eq:conservation3}
\end{align}
In fact, $\QBZM$ in \eqnref{eq:QBZM} is nothing but the charge
associated with $\jBZM^\mu$ and obviously the total baryon charge,
\begin{align}
	Q_{\text{B}} = \QB + \QBZM\,,
	\label{eq:QBtot}
\end{align}
is the conserved quantity 
regardless of the presence of vortex singularities.
As we will demonstrate in the next section, $\jBZM^\mu$ corresponds to
the baryon current carried by the vortex zeromode.  Then, the
nonconservation problem of $\jB^\mu$ in bulk is resolved with
$\jBZM^\mu$ localized on vortex cores.

Let us make one remark on $\jBZM^\mu$.
The zeromode current~\eqref{eq:jBZM} is unique only up to divergenceless
terms.  In fact, \eqnref{eq:jBZM} implies that $\jBZM^\mu$ may depend on the
extension from $\M$ to $\D$, but it is clear from \eqnref{eq:divjBZM}
that any difference caused by this would be irrelevant once the
divergence is taken.

\section{QCD axial domain-walls and vortices with magnetic fields}
\label{sec:axialdomainwall}

So far, we have seen the general theory of conservation law in systems
involving vortex singularities.  In this section we shall address two
concrete examples of such vortex configurations in QCD.
One is the $\pi^0$ domain-wall and surrounding vortex as discussed in
\secref{sec:pi}, and the other is the $\eta$ domain-wall and
surrounding vortex as discussed in \secref{sec:eta}.  In both
cases a finite baryon number is anomalously induced with coupling to
external magnetic fields.

\subsection{$\pi^0$ domain-wall and vortex}
\label{sec:pi}

\begin{figure}
 	\includegraphics[width=0.7\linewidth]{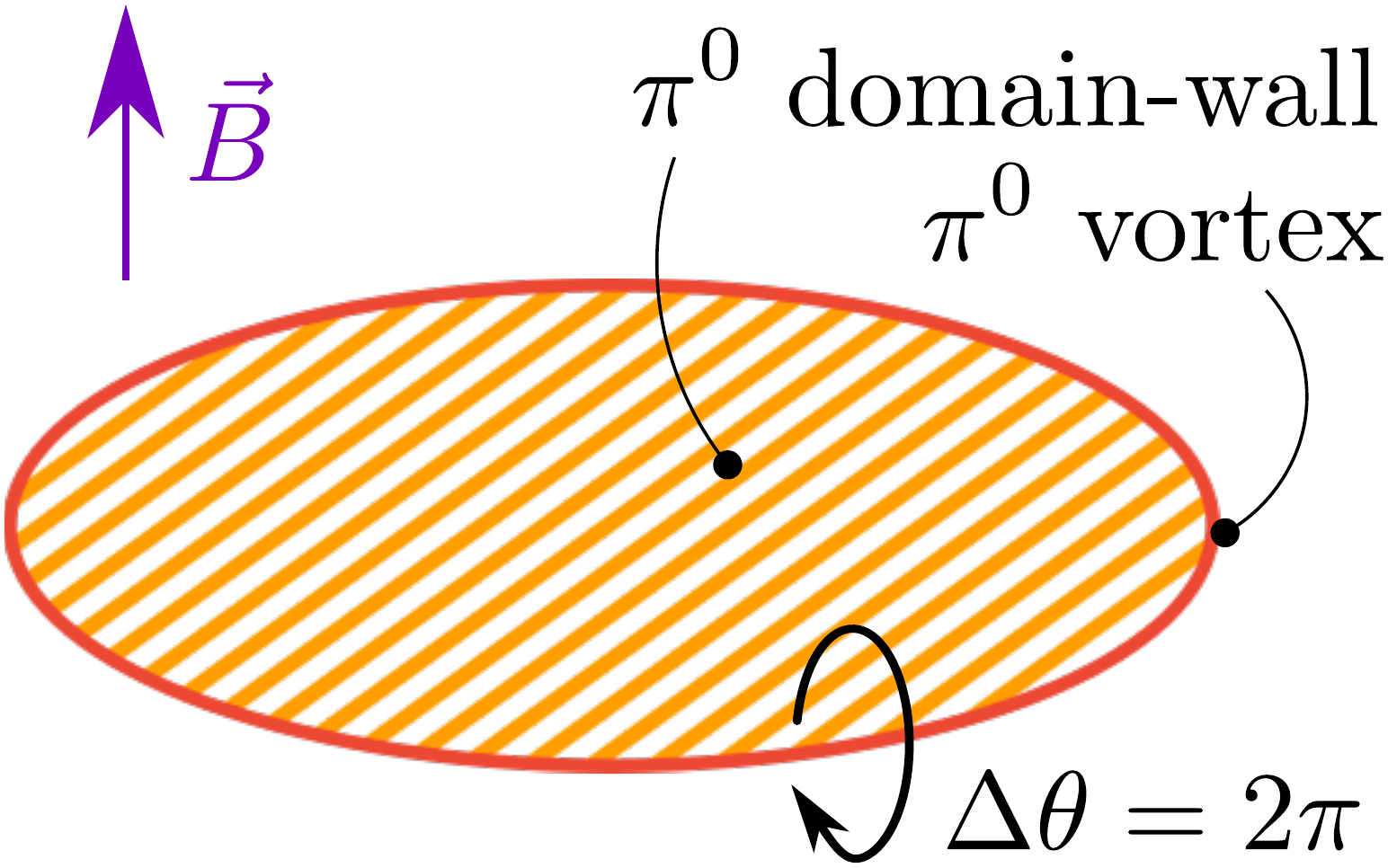}
 	\caption{
 		\label{fig:dw}
 		$\pi^0$ domain-wall and 
 		surrounding $\pi^0$ vortex string at the edge.
 		With increasing magnetic field piercing the
                 domain-wall, more baryons are attached on the
                 domain-wall according to \eqnref{eq:QBpi}.
 	}
 \end{figure}

As argued in Ref.~\cite{Brauner:2016pko}, in extreme environments with
sufficiently strong magnetic field such as the neutron star cores,
nuclear matter would form $\pi^0$ domain-wall layers.  Once the $\pi^0$
domain-wall layers develop inside the neutron star, the edge of the
$\pi^0$ domain-wall is a $\pi^0$ vortex string as
illustrated in Fig.~\ref{fig:dw}.

In the QCD vacuum the chiral
condensate spontaneously breaks chiral symmetry as
$\mathrm{SU(2)_L}\times\mathrm{SU(2)_R}\to\mathrm{SU(2)_V}$ in the
massless two-flavor case.  External
magnetic fields would explicitly break a part of chiral symmetry as
$\mathrm{SU(2)_L}\times\mathrm{SU(2)_R}\to\mathrm{U(1)_A}\times\mathrm{U(1)_V}$,
where this $\mathrm{U(1)_A}$ symmetry has nothing to do with $\eta$
meson in the anomalous sector (which will be considered in
\secref{sec:eta}) but corresponds to $\pi^0$.  Therefore, in strongly
magnetized quark matter, nonanomalous $\mathrm{U(1)_A}$ symmetry is
spontaneously broken by the chiral condensate.  In this sense quark
matter with strong magnetic fields could be regarded as a
``chiral superfluid'' which accommodates axial vortices as
topological defects.  If we consider a small but nonzero quark mass
and place a ring of $\pi^0$ vortex in the system as in
Fig.~\ref{fig:dw}, the minimal surface area surrounded by the $\pi^0$
vortex string should form a $\pi^0$ domain-wall to minimize the energy
cost by topological winding.  Let us calculate $\jB^\mu$ and
$\jBZM^\mu$ in this setup.

First, we flash theoretical descriptions following
Ref.~\cite{Son:2007ny}.  It is convenient to use the following
parametrization with $\chi$, $\theta$, $\phi$:
\begin{align}
	\Sigma
	=	\cos\chi e^{i\tau_3\theta} + i\tau_1\sin\chi e^{i\tau_3\phi}\,,
	\label{eq:param}
\end{align}
instead of conventional $\pi^0$, $\pi^\pm$, where $\tau_i$'s are Pauli
matrices in two-flavor space.
With this parametrization the nonanomalous Lagrangian~\eqref{eq:Lchi}
takes the form of
\begin{align}
	\mathcal L_\chi
		&= \frac{\fpi^2}{2} \big[
		(\p_\mu\chi)^2 + \cos^2\chi(\p_\mu\theta)^2 
		+ \sin^2\chi(\p_\mu\phi-eA_\mu)^2\notag\\
		&\qquad\quad
		-2\mpi^2(1-\cos\chi\cos\theta) \big]\,,
		\label{eq:Lchi2}
\end{align}
where $\mpi$ represents the $\pi^0$ mass.
We can easily see that 
this Lagrangian allows for a $\pi^0$ domain-wall.
If we anticipate $\cos\chi \simeq 1$ holds away
from the edge of the domain-wall for some reasons,
the Lagrangian~\eqref{eq:Lchi2} is reduced to the sine-Gordon model.
The one-dimensional classical solution in the sine-Gordon model is
well-known to be
\begin{align}
	\theta(z) = 4\arctan(e^{\mpi z})\,
	\label{eq:pi0dw}
\end{align}
which has a smooth jump of $\theta$ by $2\pi$ around $z\sim 0$.
Hence, we shall call this special solution the $\pi^0$ domain-wall.

As we already pointed out in the beginning of this section, the
$\pi^0$ domain-wall and vortex are topologically stabilized when
strong magnetic fields are imposed%
~\cite{Son:2007ny}.
In the physics language the
strong magnetic field makes $\pi^\pm$ as massive as the imposed
magnetic scale, which means $\cos\chi=1$ and $\sin\chi = 0$ are
energetically favored.  Thus, for strongly magnetized quark matter,
the sine-Gordon model is the genuine effective theory.

This $\pi^0$ domain-wall accompanies a $\pi^0$ vortex string along the
edge.  Since $\theta$ jumps by $2\pi$ at the domain-wall, $\theta$
increments by $2\pi$ along any small loop around the vortex (see
Fig.~\ref{fig:dw}).
This winding on the vortex gives rise to a singularity, i.e.\
$[\p_i,\p_j]\theta = 2\pi\delta^{(2)}(x_\perp)$, where $x_\perp$
represents a transverse coordinate perpendicular to and centered at the vortex.
It should be noted that $\sin^2\chi = 1$ on the vortex to avoid
singularity behavior of the $\theta$ kinetic term.  This implies that
$\pi^\pm$ must be present around the $\pi^0$ vortex.

Next, let us evaluate the anomaly induced baryon charge from the WZW
action.  The currents $\jB^\mu$ in \eqnref{eq:jB} and $\jBZM^\mu$ in
\eqnref{eq:jBZM} are associated with the $\pi^0$ domain-wall and the
$\pi^0$ vortex, respectively.  We can readily compute $\jB^\mu$  under
the minimization condition of the kinetic energy which concludes
$\p_\mu \phi = eA_\mu$.  Then, only the second term in \eqnref{eq:jB}
remains finite, yielding
\begin{align}
	\jB^\mu 
		= \frac{e}{8\pi^2}\epsilon^{\mu\nu\rho\sigma}
		F_{\nu\rho} \p_\sigma\theta\,.
	\label{eq:jBpi}
\end{align}
The baryon density topologically induced by the anomaly is thus
expressed as
\begin{align}
	\jB^0 
		= \frac{e}{4\pi^2} \vec B \cdot \vec\p\theta\,.
	\label{eq:jB0pi}
\end{align}
Alternatively, we can write down the integrated baryon number, $\QB$
of \eqnref{eq:QB}, even without using the energy minimization
condition.  In this case only the first term in \eqnref{eq:jB}
contributes to the spatial integration.
After all we find~\cite{Goldstone:1981kk},
\begin{align}
	\QB = \frac{1}{2\pi} \oint_{\text{vortex}} d\phi\,.
	\label{eq:QBpi}
\end{align}
We can confirm the consistency between Eqs.~\eqref{eq:jB0pi} and
\eqref{eq:QBpi} using $\p_\mu\phi = eA_\mu$ and
$\Delta \theta=2\pi$.  From this argument we understand that the
integer baryon number populates on the $\pi^0$ domain-wall for quanta
$2\pi/e$ of the piercing magnetic flux.

We can also evaluate $\jBZM^\mu$ in \eqnref{eq:jBZM}, and the
condition $\p_\mu\phi=eA_\mu$ make the final result as simple as
\begin{align}
	\jBZM^\mu
	= -\frac{1}{4\pi}\epsilon^{\mu\nu\rho\sigma}
		\p_\nu\phi\,\delta^{(2)}_{\rho\sigma}(x_\perp)\,.
	\label{eq:jBZMpi}
\end{align}
Here, $\delta^{(2)}_{\rho\sigma}(x_\perp)$ represents the
two-dimensional Dirac delta function on the $x_\rho$-$x_\sigma$ plane
centered at the vortex position.
For example, if a vortex sits along the $z$-axis,
$\delta^{(2)}_{xy}(x_\perp) = -\delta^{(2)}_{yx}(x_\perp) = \delta(x)\delta(y)$
and $\delta^{(2)}_{xx}(x_\perp)=\delta^{(2)}_{yy}(x_\perp)=0$.  Then,
the spatial volume integration of \eqnref{eq:jBZMpi} amounts to
\begin{equation}
  \QBZM = \int d^3 x\, \jBZM^0 = -\frac{1}{2\pi}\oint_{\text{vortex}}
  d\phi\,,
\end{equation}
which exactly cancels \eqnref{eq:QBpi} not to violate the conservation
law in the whole system.

We have confirmed the balance between the bulk and the zeromode
contributions to satisfy the conservation law.  It is intriguing to
argue that the balanced relation holds not only for the static case
but also for more dynamical circumstances as it should.  To see this,
let us imagine a physical setup with growing magnetic field with time.
Such time-dependent magnetic field should generate an eddy electric
field via Faraday's law, and \eqnref{eq:jBpi} leads to%
~\cite{Yamamoto:2015maz}
\begin{align}
	\vec j_{\BA,\bu} 
		= -\frac{e}{8\pi^2} \vec E \times \vec\p\theta\,.
	\label{eq:jBipi}
\end{align}
Because $\vec{E}$ is along the polar angular direction and $\theta$
has spatial variation perpendicular to the $\pi^0$ domain-wall, the
current is directed along the radial coordinate on the $\pi^0$
domain-wall, i.e., either inward to or outward from the $\pi^0$ vortex
string where the baryon charge is absorbed or emitted.

\subsection{$\eta$ domain-wall and vortex}
\label{sec:eta}

If anomalous $\mathrm{U(1)_A}$ symmetry is effectively restored in
extreme conditions to suppress instanton excitations, we can
anticipate the spontaneous breaking of $\mathrm{U(1)_A}$ and
associated vortices with respect to $\eta$ mesons.

The most well-known example of extreme conditions to accommodate
$\eta$ vortices is the color-flavor locked state of quark matter at
high density%
~\cite{Alford:2003fq}.
Instantons are expected to be Debye screened by the
density effect and chiral symmetry including $\mathrm{U(1)_A}$ is
spontaneously broken by the diquark condensates there.  In this state
the phase, $\phiA$, of $\Sigma = |\Sigma|\,e^{i\phiA}$, is the
$\eta_0$ meson (which we simply call $\eta$ in this paper) and becomes
an Nambu-Goldstone boson.  The effective Lagrangian for $\phiA$ is
given by%
~\cite{Schafer:2002ty, Son:2000fh}
\begin{align}
	\mathcal L_\eta 
	= \frac{f_\eta}{2} \bbr{(\p_0\phiA)^2-u^2(\vec\p\phiA)^2 
	-2m_\eta^2(1-\cos \phiA)}\,,
	\label{eq:etakineticterm}
\end{align}
where $f_\eta$ and $m_\eta$ are the decay constant and the mass of the
$\eta$ meson.
We note that $u^2$ represents the speed of $\eta$ and is not necessarily
the unity in a medium which breaks Lorentz symmetry.

The Lagrangian~\eqref{eq:etakineticterm} is of the sine-Gordon type
and admits an $\eta$ domain-wall solution characterized by a jump of
$\phiA$ by $2\pi$.  In the same way as the $\pi^0$ domain-wall, this
$\eta$ domain-wall also accompanies the $\eta$ vortex string at the
edge leading to $[\p_i,\p_j]\phiA = 2\pi\delta^{(2)}(x_\perp)$ at the
vortex core.

The rest of the discussions goes parallel to the previous subsection.
We can evaluate $\jB^\mu$ and $\jBZM^\mu$ associated with the $\eta$
domain-wall and the $\eta$ vortex, respectively.  Then, a
straightforward computation results in the following
expression~\cite{Son:2004tq},
\begin{align}
	\jB^\mu
		= \frac{e}{24\pi^2} \epsilon^{\mu\nu\rho\sigma} 
		F_{\nu\rho}\p_\sigma\phiA\,,
	\label{eq:jBeta}
\end{align}
which is only factor $1/3$ different from \eqnref{eq:jBpi}.  
Actually, from the microscopic point of view, the difference between $\pi^0$ and
$\eta$ is only a relative sign between $\bar{u}u$ and $\bar{d}d$. 
We can explain this factor difference between $\frac{2}{3}e-(-\frac{1}{3}e)=e$ and
$\frac{2}{3}e+(-\frac{1}{3}e)=\frac{1}{3}e$.
We find the zeromode contribution as
\begin{align}
	\jBZM^\mu
		= -\frac{e}{12\pi}\epsilon^{\mu\nu\rho\sigma}
		A_\nu\, \delta^{(2)}_{\rho\sigma}(x_\perp)\,.
	\label{eq:jBZMeta}
\end{align}
Using $\p_\mu\phi=eA_\mu$, we see that \eqnref{eq:jBZMeta} is only
factor $1/3$ different from \eqnref{eq:jBZMpi}.

Before closing this subsection, let us point out an interesting
observation which motivates our analysis in the next section.  If we
take the divergence of the zeromode current, we will have a relation
analogous to the anomaly, i.e.\ we immediately get the following
from \eqnref{eq:jBZMeta},
\begin{align}
	\p_\mu \jBZM^\mu
	= -\frac{e}{24\pi}\epsilon^{\mu\nu\rho\sigma}
		F_{\mu\nu}\,\delta^{(2)}_{\rho\sigma}(x_\perp)\,.
	\label{eq:divjZMeta}
\end{align}
In fact, along the vortex as dictated by
$\delta^{(2)}_{\rho\sigma}(x_\perp)$, the above is nothing but the
gauge anomaly relation for (1+1)-dimensional chiral fermions with a
total charge $\frac{2}{3}e-\frac{1}{3}e=\frac{1}{3}e$.
Such formal similarity is not accidental but is attributed to physical
contents of the vortex in terms of chiral $u$- and $d$-quarks, as we
discuss in the next section.

\section{Anomaly Inflow}
\label{sec:inflow}

It may look nontrivial how our arguments in the previous section,
based on the gauged WZW action,
are related to the ``zeromode'' contributions from the fermionic sector.
To fill in this gap, we shall make a brief review on
Callan-Harvey's mechanism of the anomaly inflow 
on axion vortices following Refs.~\cite{Callan:1984sa}.

We introduce axion $\axion(x)$ coupled to fermions as
\begin{align}
	S = \int d^4x\,
		\bar\psi \bbr{i\gamma^\mu(\p_\mu + i e A_\mu) - G\Phi} \psi
\end{align}
with $\Phi=|\Phi| e^{i\gamma_5\axion}$~\cite{Peccei:1977hh}.  The
coupling constant $G$ will turn out to be irrelevant in what follows
below.  Here, we consider only one flavor for simplicity but
generalization is straightforward.  Suppose that the symmetry with
respect to the $\axion$ rotation (that is, Peccei-Quinn symmetry) is
spontaneously broken to set $|\Phi| = 1$.  The current expectation
value is expressed by the fermion propagator as
\begin{align}
	j^\mu(x)
	= - \tr[\gamma^\mu P(x,y\to x)]\,,
	\label{eq:formulaofcurrent}
\end{align}
where the concrete form of the propagator is
$P=i/(i \gamma^\mu \p_\mu - G\Phi)$.  In the long wavelength limit,
$G\to\infty$, we can expand the propagator in terms of $1/G$ and a
nonzero contribution remains as
\begin{align}
	j^\mu(x)
	&= -2e \tr[\gamma^\mu\gamma^\nu\gamma^\rho\gamma^\sigma\gamma_5]
	\p_\nu A_\rho \p_\sigma\axion
	\!\int\!\! \frac{d^4 p}{(2\pi)^4}\frac{G^2}{(p^2\!-G^2)^3}\notag\\
	&= \frac{e}{8\pi^2}\epsilon^{\mu\nu\rho\sigma}
	F_{\nu\rho}\,\p_\sigma\axion\,,
	\label{eq:axion_j}
\end{align}
whose divergence reads,
\begin{align}
	\p_\mu j^\mu
	= \frac{e}{8\pi^2}\epsilon^{\mu\nu\rho\sigma}
	F_{\nu\rho}\,\p_\mu\p_\sigma\axion\,.
\end{align}
One might think that this divergence vanishes, but this is not the
case when an ``axion vortex string'' is present, which causes a
singularity as
$[\p_\mu, \p_\sigma]\axion = 2\pi\delta^{(2)}_{\mu\sigma}(x_\perp)$,
yielding,
\begin{align}
	\p_\mu j^\mu
	= \frac{e}{8\pi} \epsilon^{\mu\nu\rho\sigma}
        F_{\mu\nu}\, \delta^{(2)}_{\rho\sigma}(x_\perp)\,,
\label{eq:axion_divj}
\end{align}
which apparently indicates violation of the gauge symmetry.

The key ingredient to resolve this apparent puzzle comes from a
zeromode in the fermionic sector, which emerges in the presence of the
axion vortex%
~\cite{Witten:1984eb,*Jackiw:1981ee}.
If the Dirac operator, $(i\gamma^\mu\p_\mu-G\Phi)$, has
a zero eigenvalue, we cannot take the inversion in the evaluation of
\eqnref{eq:formulaofcurrent} and should separately calculate the
zeromode contribution,
\begin{align}
	j^\mu_\zm
	= \bar\psi_\zm\gamma^\mu\psi_\zm\,.
\end{align}
Here, $\psi_\zm$ represents the zeromode solution of the Dirac
equation.  To help our intuitive understanding, let us concretely
specify the vortex configuration.  Below we make use of the
cylindrical coordinates $(r,z,\varphi)$ and place an axial vortex,
$\axion(\varphi)=+\varphi$, along the $z$-axis at $r=0$.  Then, the
zeromode equations read,
\begin{subequations}
\begin{align}
	&\bbr{ i\gamma^\alpha\p_\alpha 
		+ i\gamma^1(\cos\varphi+\gamma^1\gamma^2\sin\varphi)\p_r}
		\psi_{\zm,\mathrm L} \notag\\
	&\qquad\qquad\qquad\qquad\qquad\qquad\quad
                = G|\Phi| e^{i\axion} \psi_{\zm,\mathrm R}\,,
                \label{eq:ZMeq1}\\
	&\bbr{ i\gamma^\alpha\p_\alpha 
		+ i\gamma^1(\cos\varphi+\gamma^1\gamma^2\sin\varphi)\p_r}
		\psi_{\zm,\mathrm R} \notag\\
	&\qquad\qquad\qquad\qquad\qquad\qquad\quad
		= G|\Phi| e^{-i\axion} \psi_{\zm,\mathrm L}\,,
                \label{eq:ZMeq2}
\end{align}
\end{subequations}
where 1 and 2 of $\gamma^\mu$ refer to the transverse directions
perpendicular to the $z$ or 3 direction.  The index $\alpha$ runs over
0 and 3 only.  We can solve these equations to get the zeromode
solutions as
\begin{align}
	\psi_{\zm,\mathrm L}
	&= \exp\bbr{-\int_0^r d r'\, G|\Phi|(r')} \xi(t,z)
	\label{eq:axion_zm}
\end{align}
and $\psi_{\zm,\mathrm R} = -i\gamma^1 \psi_{\zm,\mathrm L}$.  Here,
$\xi(t,z)$ is a solution of the free Dirac equation, i.e.
\begin{align}
	\gamma^\alpha\p_\alpha \xi = 0
\end{align}
with right-handed ``chirality'' defined by
$\gamma^0\gamma^3\xi = +\xi$, which implies $(\p_0+\p_z)\xi=0$.
The exponential factor in \eqnref{eq:axion_zm} shows that
this solution is localized near the axion vortex string and
approaches $\delta^{(2)}(x_\perp)$ in the limit $G\to\infty$.
Then, the (1+1)-dimensional effective action for the zeromode on the
axial vortex takes the following form,
\begin{align}
	S_\zm
	= \int dt\, dz\,
		\bar\xi \bbr{i\gamma^\alpha (\p_\alpha+i eA_\alpha)} \xi\,,
	\label{eq:axion_Szm}
\end{align}
where we retrieved the gauge field in the covariant derivative.  This
(1+1)-dimensional theory exhibits the gauge anomaly to conclude
$\p_\alpha j_\zm^\alpha=-(eE_z/2\pi)\delta^{(2)}(x_\perp)$.  It is a
straightforward exercise to reexpress this result in a covariant
manner, which finally yields,
\begin{align}
	\p_\mu j_\zm^\mu
	= -\frac{e}{8\pi}\epsilon^{\mu\nu\rho\sigma}
        F_{\mu\nu}\,\delta^{(2)}_{\rho\sigma}(x_\perp)\,.
	\label{eq:axion_divjZM}
\end{align}
In this way the apparent violation of the gauge symmetry in
\eqnref{eq:axion_divj} is precisely canceled by the zeromode
contribution of \eqnref{eq:axion_divjZM}.  This means that, if we take
account of both the ``inflow'' of \eqnref{eq:axion_divj} in bulk and
the ``leakage'' of \eqnref{eq:axion_divjZM} from the edge, 
the whole system conserves the charge as it should.

Let us then reinterpret the arguments above in the language of the
effective action.  As we shall see immediately, the effective action
of the system involving axion vortices is given by the following
five-dimensional form:
\begin{align}
	S_{\text{axion}} = -\frac{e^2}{8\pi^2} \int_{\D} d^5x\,
		\epsilon^{\MU\NU\RHO\SIGMA\TAU}
		\, F_{\MU\NU} F_{\RHO\SIGMA}\,\p_\TAU \axion\,.
\end{align}
To see this,
decompose this action into the four- and the five-dimensional
pieces, which is reminiscent of \eqnref{eq:wzw}.
That is,
\begin{align}
  S_{\text{axion}}
  = -\int_{\M} d^4x\, eA_\mu j^\mu
    -\int_{\D} d^5x\, eA_\MU \mathfrak{j}^\MU
	\label{eq:Saxion}
\end{align}
with the four-dimensional current $j^\mu$ is given by
\eqnref{eq:axion_j} and the five-dimensional current
$\mathfrak{j}^\MU$ is defined by
\begin{align}
	 \mathfrak{j}^\MU = \frac{e}{8\pi^2}
		\epsilon^{\MU\NU\RHO\SIGMA\TAU}
		F_{\NU\RHO}
		\, \p_\SIGMA\p_\TAU \axion\,.
\end{align}
The gauge invariance of the whole action assures that
the five-dimensional part in \eqnref{eq:Saxion} cancels 
the gauge variance of the other term 
in the presence of vortex singularity.
This five dimensional term thus reproduces the gauge anomaly of 
the chiral zeromode given by \eqnref{eq:axion_Szm}%
~\cite{stone1991edge}.
This cancellation assures that the zeromode current,
defined in the same manner to \eqnref{eq:jBZM},
gives the divergence identical to \eqnref{eq:axion_divjZM}.

Now the analogy to the case with the QCD axial vortices is clear;
the five-dimensional action in \eqnref{eq:wzw},
derived from \eqnref{eq:witten0},
gives the consistent effective action inclusive of
the anomalous action of fermionic zeromodes.
Indeed, for the systems we considered in the \secref{sec:axialdomainwall}
we could explicitly check the consistency about $\jBZM^\mu$ from the
zeromode construction and that from the effective action approach.
For the $\eta$ vortex, \eqnref{eq:divjZMeta} is immediately reproduced
from \eqnref{eq:axion_divjZM}.
Interested readers can further consult \appref{app:pi} for the microscopic
derivation of the zeromode current.

\section{Conclusions}
\label{sec:conclusion}

We considered a general problem of gauge invariance of the WZW action
for singular configurations which make the order of derivatives not
commutative.  This is not an academic subject, but provides us with an
insight to understand how the topologically induced charges should be
conserved.

As concrete examples we discussed the $\pi^0$ domain-wall and vortex
and the $\eta$ domain-wall and vortex.  In the presence of external
magnetic field the baryon and the electric charges are attached on the
domain-walls due to the anomaly coupling.  We explicitly checked that
such induced charges are precisely canceled by singular contributions
from the domain-wall edge, that is, the surrounding vortex ring or the
vorton.

Interestingly, this cancellation mechanism has the same theoretical
structure as Callan-Harvey's mechanism of the anomaly inflow for the
axion vortex;  the anomaly on the domain-wall in bulk is canceled by
another anomaly in the (1+1)-dimensional fermionic sector on the
vortex.

In the present work we limited ourselves to the two-flavor case and
the three-flavor extension would be an intriguing future problem.
Then, it is known that the genuine ground state of dense quark matter
is definitely the CFL phase if the baryon density is large enough.
Since the CFL phase is a superfluid, it accommodates topologically
stable vortices.  Thanks to color and flavor degrees of freedom, there
may appear many different types of vortices.  
Some changes in the linkage number of various vortex entanglements could cause new
topological effects, in the same way as the chiral magnetic effect
associated with the linkage number change in terms of the magnetic
fluxes~\cite{Hirono:2016jps}.
A possible realization of the vortex linkage is found in the  CFL phase
with kaon condensation which supports vortons, i.e.,
topologically stable vortex
rings~\cite{Buckley:2002ur,Kaplan:2001hh,Buckley:2002mx}.
Mixed condensates of $K^0$ and $K^+$~\cite{Kaplan:2001qk} lead to
both $K^0$ and $K^+$ vortices.
Since a $K^+$ vortex carries a magnetic flux, 
a non-trivial linkage between the $K^0$ and $K^+$ vortices would
correspond to a situation that a magnetic flux pierces a
vorton~\cite{Kaplan:2001hh}, as we considered in this work.
For the microscopic description of the charge density and
currents in such systems, the extra terms involving the derivative commutator that we
found in this work would play an essential role.  The anomaly inflow
in the context of dense QCD with vortices would certainly deserve
further investigations.

\begin{acknowledgments}
The authors thank
Naoki~Yamamoto
and
Ho-Ung~Yee
for useful discussions and conversations.
This work was supported by Japan Society for the Promotion of Science
(JSPS) KAKENHI Grant No.\ 15H03652 and 15K13479.
\end{acknowledgments}        

\appendix

\section{Microscopic derivation of the zeromode current for the $\pi^0$ vortex}
\label{app:pi}

We begin with the fermionic action for quarks,
\begin{align}
	S = \int d^4x \bbr{
	i \bar q\gamma^\mu\p_\mu q
	-G (\bar q_{\mathrm L}\Sigma q_{\mathrm R}
+\bar q_{\mathrm R}\Sigma^\dagger q_{\mathrm L}) }\,,
\end{align}
where $q$ represents the quark field 
and $G$ is the Yukawa coupling constant, which we will send to
infinity in the end of the calculation.
In cylindrical coordinates $(r,z,\varphi)$, a $\pi^0$ vortex lying
along the $z$-axis is parametrized by
\begin{align}
	\Sigma = \cos\chi(r) e^{i\tau_3\theta(\varphi)}
	+ i\tau_1 \sin\chi(r) e^{i\tau_3\phi(t,z)}
\end{align}
with $\theta(\varphi) = +\varphi$ and the boundary condition,
$\cos\chi \to 1$ for $r \to \infty$, should be imposed.  The zeromode
equations corresponding to Eqs.~\eqref{eq:ZMeq1} and \eqref{eq:ZMeq2}
are
\begin{subequations}
\begin{align}
	&\bbr{ i\gamma^\alpha\p_\alpha 
		+ i\gamma^1(\cos\varphi+\gamma^1\gamma^2\sin\varphi)\p_r}q_{\mathrm L}\notag\\
		&\qquad\qquad\qquad
		= G (\cos\chi e^{i\tau_3\varphi} + i\sin\chi e^{i\tau_3\phi})
		q_{\mathrm R}\,,\\
	&\bbr{ i\gamma^\alpha\p_\alpha 
		+ i\gamma^1(\cos\varphi+\gamma^1\gamma^2\sin\varphi)\p_r}q_{\mathrm R}\notag\\
		&\qquad\qquad\qquad
		= G (\cos\chi e^{i\tau_3\varphi} + i\sin\chi e^{i\tau_3\phi})
		q_{\mathrm L}\,,
\end{align}
\end{subequations}
where $\alpha = 0,\,3$.  We can write down an explicit solution for
sufficiently large $G$ as
\begin{align}
	q_{\mathrm L} 
	&= \exp\bbr{-\int_0^r d r'\, G\cos\chi(r')} \xi(t,z)
	\label{qLpi}
\end{align}
and $q_{\mathrm R} = -i\gamma^1 q_{\mathrm L}$.  Here, $\xi(t,z)$
satisfies not the free Dirac equation but the equation with $\phi$
background, that is,
\begin{align}
	\pbr{ i\gamma^\alpha\p_\alpha
	-G\tau_1 \gamma^1 e^{i\tau_3\phi} }\xi=0\,.
\end{align}
We note that the chirality should be flipped depending on the quark
flavors, i.e., $\tau_3\gamma^0\gamma^3\xi = +\xi$.
The exponential factor is reduced to $\delta^{(2)}(x_\perp)$ in the
limit of $G\to\infty$.
To sort expressions out, it would be convenient to introduce a new set
of Dirac matrices,
\begin{align*}
	\Gamma^0 = i\tau_1\gamma^2\gamma^3\,, \qquad
	\Gamma^3 = i\tau_1\gamma^2\gamma^0\,, \qquad
	\Gamma_5 = \tau_3\,.
\end{align*}
Under the constraint, $\tau_3\gamma^0\gamma^3=+1$, we can prove the
Clifford algebra, $\{\Gamma^\alpha,\Gamma^\beta\} = 2g^{\alpha\beta}$,
$\{\Gamma^\alpha,\Gamma_5\} = 0$, and $(\Gamma_5)^2 = 1$.  Then, the
(1+1)-dimensional effective action for the zeromode on the $\pi^0$
vortex takes the following form,
\begin{align}
	S_\zm
	= \int dt dz\,
	{\bar\xi} 
	\pbr{ i\Gamma^\alpha\p_\alpha - G e^{i\Gamma_5\phi}} \xi
	\label{2dtheorypi}
\end{align}
with ${\bar\xi} = \xi^\dagger\Gamma^0$.
The most efficient method to infer the current from this effective
action is to rewrite this into the bosonized form in terms of boson
field $\sigma$, namely, $S_\zm\to S'_\zm$ with
\begin{align}
	S'_\zm
	= \Nc\int dt dz\,
		\bbr{\frac1{8\pi}\p_\mu \sigma\p^\mu \sigma -G\cos(\sigma-\phi)}\,,
\end{align}
where $\Nc$ is the color number of quarks%
~\cite{Goldstone:1981kk}.
The (1+1)-dimensional quark current is given by $j^\alpha
=  (\Nc/2\pi)\epsilon^{\alpha\beta}\p_\beta \sigma$.  From the minimum
of the potential $G\cos(\sigma-\phi)$, we should plug $\sigma=\phi$
into this expression.  The overall factor $\Nc$ is eliminated by the
conversion from the quark current to the baryon current of our
interest, and taking account of $\delta^{(2)}(x_\perp)$ for the
current in the original (3+1)-dimensional theory, we finally arrive at
\begin{align}
	\jBZM^\mu 
	= -\frac{1}{4\pi}\epsilon^{\mu\nu\rho\sigma}
        \p_\nu \phi\,\delta_{\rho\sigma}^{(2)}(x_\perp)\,,
\end{align}
which is nothing but \eqnref{eq:jBZMpi}.

\bibliography{inflow}
\bibliographystyle{apsrev4-1}
\end{document}